# Residual Kriging for Regional-Scale Canopy Height Mapping: Insights into GEDI-Induced Anisotropies and Sparse Sampling

Kamel Lahssini
*UMR TETIS*
*INRAE*
Montpellier, France
0000-0002-5359-8718

Guerric le Maire
*UMR Eco&Sols*
*CIRAD*
Montpellier, France
0000-0002-5227-958X

Nicolas Baghdadi
*UMR TETIS*
*INRAE*
Montpellier, France
0000-0002-9461-4120

Ibrahim Fayad
*LSCE/IPSL*
*Université Paris-Saclay*
Gif-sur-Yvette, France

*Abstract*—Quantifying aboveground biomass (AGB) is essential in the context of global climate change. Canopy height, which is related to AGB, can be mapped using machine learning models trained with multi-source spatial data and Global Ecosystem Dynamics Investigation (GEDI) measurements. In this study, a comparative analysis of canopy height estimates derived from two models is presented: a U-Net deep learning model (*CHNET*) and a Random Forest algorithm (*RFH*). Both models were trained using GEDI lidar data and utilized multi-source inputs, including optical, radar, and environmental data. While *CHNET* can leverage its convolutional architecture to account for spatial correlations, we observed that it does not fully incorporate all the spatial autocorrelation present in GEDI canopy height measurements. By conducting a spatial analysis of the models' residuals, we also identified that GEDI data acquisition parameters, particularly the variability in laser beam energy combined with the azimuthal directions of the observation tracks, introduce spatial inconsistencies in the measurements in the form of periodic patterns. To address these anisotropies, we considered exclusively GEDI power beams, and we conducted our spatial autocorrelation analysis in the GEDI track azimuthal direction. Next, we employed the residual kriging (RK) spatial interpolation technique to account for the spatial autocorrelation of canopy heights and improve the accuracies of *CHNET* and *RFH* estimates. Adding RK corrections improved the performance of both *CHNET* and *RFH*, with more substantial gains observed for *RFH*. The corrections appeared to be localized around the GEDI sample points and the density of usable GEDI information is therefore an important factor in the effectiveness of spatial interpolation. Furthermore, our findings reveal that a Random Forest model combined with spatial interpolation can deliver performance comparable to that of a U-Net model alone.

## I. Introduction

Tropical forests are among the most important natural ecosystems on Earth as they play a crucial role in regulating the global climate. These forests are not only biodiversity centres but also act as significant carbon sinks, absorbing large amounts of carbon dioxide from the atmosphere [1], [2], [3]. The preservation of the standing aboveground biomass (AGB) within these ecosystems is key in the context of global warming and climate change. By storing carbon, tropical forests mitigate the greenhouse effect and thus help to stabilize global temperatures. However, the ongoing deforestation and degradation of these forests threaten their ability to function as carbon sinks, making it imperative to monitor accurately AGB levels.

To effectively quantify AGB, studies commonly rely on allometric equations that link the structural characteristics of a forest, such as tree height, diameter at breast height, and wood density, to its biomass [4], [5]. These relationships can be applied either at the individual tree level or at stand level. Precise estimates of AGB can be obtained from in situ measurements but they are limited to small and accessible areas. Canopy height is of paramount importance in AGB estimation models. This variable can be estimated using remote sensing techniques. Accurate canopy height measurements are essential for deriving reliable AGB estimates through allometric models that rely on canopy height. Over the past decade, canopy height estimates have been produced using various remote sensing data sources and methodologies [6], [7], [8], [9], [10]. However, these estimates often exhibit substantial uncertainty, especially in dense and complex tropical environments where the retrieval of accurate heights is more challenging [11], [12]. Reported estimation error values for canopy height products derived from remote sensing can vary widely depending on the region and the datasets used. For example, a recent validation study of three widely used global canopy height maps [8], [9], [10] reported root mean squared errors (RMSEs) ranging from 9 to 18 m, with a consistent tendency to underestimate tall canopies [13]. These levels of error are well above the precision thresholds required for applications such as the United Nations' REDD+ program, which emphasizes the importance of precision in AGB estimates and recommends a relative error of 20% [14], [15]. Therefore, improving the accuracies of large-scale canopy height estimates is needed to meet these international standards and to ensure the effectiveness of global forest conservation efforts.

In this context, Light Detection and Ranging (lidar) systems have emerged as powerful tools for characterizing vegetation profiles and structural parameters. Airborne lidar scanning (ALS) systems provide high point densities and fine spatial resolution (often sub-meter), making them well suited for accurate canopy height mapping at local and regional scales [16]. However, ALS data are typically limited in spatial extent due to the cost and logistical complexity of acquisition campaigns. They also tend to cover selected areas, mostly in developed regions such as Europe and North America [17]. In contrast, spaceborne lidar sensors offer the advantages of consistent acquisition strategies and global coverage, which are advantageous for producing canopy height estimates over large extents. Their measurements can serve as reference data in conjunction with other remote sensing sources to generate regional and global canopy height maps at finer resolutions [8], [9], [12]. Nonetheless, the accuracy of these global maps remains limited and the prediction of canopy height heterogeneity is still greatly underestimated [13]. Moreover, spaceborne lidar systems face several challenges linked with their acquisition process. Indeed, their sampling strategy results in sparsely and unevenly distributed footprints, and products derived directly from spaceborne lidar measurements often have coarse spatial resolution due to the need for spatial aggregation to ensure sufficient sampling density [18]. Their high operational altitudes also increase their sensitivity to environmental conditions and can limit their effectiveness [19], [20], [21]. The Global Ecosystem Dynamics Investigation (GEDI), a system specifically designed to measure vegetation structure on a global scale, represents the latest advancement in spaceborne lidar technology [22]. As a full-waveform system, GEDI captures waveforms that directly represent the vegetation's vertical structure. From these raw waveforms, various descriptive metrics can be extracted to characterize canopy height. The simplest and most straightforward method consists in using a single metric for direct canopy height estimation.

Since canopy height estimation is only performed at GEDI footprint locations given the sparse nature of GEDI information, it is necessary to combine these point-based estimates with continuous bidimensional remote sensing data to generate comprehensive and continuous canopy height maps. GEDI data are used as reference canopy heights to train and validate prediction models that usually integrate optical and radar information. In this perspective, statistical and machine learning models, such as stepwise regression or Random Forest, have been widely employed for accurate canopy height mapping at various scales [8], [23]. Recent advancements in deep learning, particularly Convolutional Neural Networks (CNNs), have significantly improved the integration of complementary data sources [24]. In the context of canopy height mapping from multiple data sources, CNNs are able to integrate spatial and textural information at both local and global scales [25], [26], [27]. Consequently, convolutional architectures have been effective in combining GEDI metrics with optical and radar images to produce extensive and continuous canopy height maps [9], [28]. The incorporation of additional ancillary environmental parameters related to canopy structure can also further enhance the accuracy of canopy height estimates [12].

Estimating forest parameters and especially canopy height from multiple remote sensing data is challenging in tropical biomes, which are characterized by dense vegetation and tall canopies. A major issue affecting canopy height estimates in these regions is the saturation of sensors in areas with high AGB levels [29]. For example, even though multispectral data do not measure height directly, they rely on spectral indices that correlate with canopy structure. These indices often saturate in dense tropical forests, thus reducing their sensitivity to height differences above a certain threshold [7], [30]. Similarly, radar signals, which are sensitive to vegetation structure through signal backscatter, also face reduced penetration in dense canopies. Even long-wavelength radar systems such as L-band can be affected when AGB exceeds certain levels, leading to a reduced sensitivity to vertical structure [31]. Regarding GEDI, a key factor impacting height estimates is the sensor's ability to penetrate through dense vegetation to accurately capture the whole vertical structure from the canopy top to the ground [32]. On the contrary, in areas with sparse or low vegetation, GEDI may also overestimate canopy height by up to several meters, likely because of signal noise, terrain slope, or bad identification of the ground return [33], [34].

Canopy height is a continuous variable influenced by various environmental factors, and traditional approaches such as linear regressions or Random Forest algorithms often lack the inherent incorporation of spatial correlation in their design [35]. As a result, these methods may not fully explain the spatial autocorrelation of canopy heights and leave some unexplained variance which could be linked to forest endogenous processes. Regarding CNNs, despite being able to extract spatial and textural features, the question of their ability to fully exploit the spatial information contained in raw input data sources and reference data is still an open research question. In this regard, spatial interpolation techniques can be employed alongside canopy height regression models to take into account the spatial autocorrelation of the data and refine canopy height estimates. In particular, residual kriging (RK) is a geostatistical technique that relies on the intrinsic stochastic properties of a dataset [36]. Contrary to deterministic interpolation techniques, which use mathematical functions to calculate the values at unsampled locations based on the degree of similarity with respect to known points at sampled locations, RK uses both analytical and statistical methods to predict unknown values based on the spatial autocorrelation of the data [37]. RK is widely employed in climatic, water, and soil applications [36], [37], [38], yet its use in forestry and particularly for estimating canopy height remains relatively rare [39], [40]. By integrating RK with canopy height regression models, it is possible to better capture the spatial patterns and variations in canopy height. In the context of canopy height mapping with sparse GEDI reference data, RK consists in an ordinary kriging (OK) procedure applied to the residuals of GEDI-based regression models. The resulting interpolated residuals, known as kriged residuals, can then be added to the regression results. This technique allows for the adjustment of canopy height estimates by accounting for the spatial structure of the residual errors that are not explained by the regression model. To accurately describe this spatial structure, it is essential to consider the specificities of the GEDI sensor, as its acquisition process may introduce spatial anisotropies that do not represent the true spatial autocorrelation of canopy heights [41].

In this study, we conduct a spatial analysis of two regional-scale canopy height maps of French Guiana, both generated from the same multi-source remote sensing inputs and trained on GEDI reference data using two different modelling approaches: a U-Net deep learning model (*CHNET*) and a Random Forest algorithm (*RFH*). While convolutional networks like U-Net are designed to capture spatial patterns, their ability to fully exploit the spatial autocorrelation of reference data such as GEDI remains uncertain. In contrast, Random Forest is a non-spatial model, which makes it a useful baseline for comparison.

At the same time, the GEDI instrument itself may introduce spatial artifacts into canopy height measurements due to its acquisition configuration, particularly because of the variations in laser beam power and the azimuthal ground track directions. These effects could introduce spatial anisotropies in the measurements that do not reflect the true forest spatial structure. To address these concerns and improve canopy height estimations, we explore the integration of residual kriging (RK) as a spatial interpolation step applied to the residuals of both *CHNET* and *RFH*. Our research is structured around the following key questions:

1. To what extent do the *CHNET* and *RFH* models capture the spatial autocorrelation of canopy height derived from GEDI measurements?
2. Do GEDI acquisition parameters (for example beam energy and track azimuth) introduce spatial anisotropies that affect model residuals?
3. Can RK effectively correct for these sensor-induced spatial effects and improve canopy height predictions?
4. How does the density of usable GEDI data influence the effectiveness of spatial interpolation?
5. Is RK more beneficial when applied to *RFH* (a non-spatial method) than to *CHNET* (a model that inherently captures spatial patterns)?

By structuring our research around these questions, we aim to clarify the role of spatial information in canopy height modeling, assess the utility of RK as a post-processing step, and provide insights for future large-scale canopy height mapping efforts using GEDI and similar data sources.

## II. Materials and Methods

### A. Study Area

French Guiana, an overseas territory of France, is situated within the Amazon biome on the northern coast of South America. It covers an area of 83,534 km², with more than 80,000 km² of forests [42]. The primary forest type is mature old-growth tropical rainforest, while some areas contain secondary forests [43]. The coastal zones also include savannas and mangroves, but rainforest covers over 90% of the territory. Timber extraction and agricultural activities are largely concentrated in the sub-coastal regions close to major towns and along the main roads [44]. The region's terrain is mostly flat, with ground elevations rarely surpassing 200 m, though some small hills and mountains can be found in the landscape [43]. Approximately 70% of the slopes are less than 5° [42]. French Guiana has a hot, tropical climate, classified as tropical rainforest (Af) under the Köppen climate classification [45]. The average annual temperature is around 26°C, while rainfall varies significantly, reaching up to 4,000 mm per year in the northeast and about 2,000 mm in the southern and western areas [44]. The combination of climatic, geological, and geomorphological factors creates favorable conditions for the development of diverse forest structures across the region. In French Guiana's tropical forests, canopy heights typically range from 20 to 40 m, with some emergent trees reaching as tall as 60 m [46]. AGB varies widely, generally between 150 Mg/ha and over 600 Mg/ha, with higher values found in mature and undisturbed forests [42].

### B. Datasets

#### 1) GEDI Reference Data

The Global Ecosystem Dynamics Investigation (GEDI) instrument is a spaceborne lidar system mounted on the International Space Station (ISS), specifically designed to characterize the structure and dynamics of forest ecosystems. It is a joint mission between NASA and the University of Maryland that has acquired and processed data for the period between March 2019 and March 2023. The system utilizes three 1064 nm lasers, which emit 242 pulses per second, to produce energy return waveforms (L1B product) and waveform-derived height metrics (L2A product) within circular footprints of 25 meters in diameter. Consequently, the data is sparsely and unevenly distributed, with GEDI covering only about 4% of Earth's surface [22].

The ISS orbits Earth at a Low Earth Orbit (LEO), at an altitude of approximately 400 km and an inclination of 51.6° relative to the equator. This inclination allows GEDI to cover latitudes between approximately 51.6°N and 51.6°S. The ISS follows a near-circular prograde orbit, meaning it travels in the same direction as Earth's rotation (from the west to the east), completing one full revolution around Earth every 90 minutes. Due to its inclined orbit, the ISS's projected ground track exhibits a sinusoidal shape and therefore the subsequent GEDI's ground tracks show variability in azimuthal direction. Specifically, in regions near the equator, such as French Guiana, the ground tracks are distributed in two main azimuthal directions (Figure 1a). In the context of this study, these two configurations are referred to as northward pass (NWD) and southward pass (SWD). A northward pass occurs when the ISS ground track moves from lower to higher latitudes (i.e., south to north), while a southward pass occurs when the ISS ground track moves from higher to lower latitudes (i.e., north to south). In French Guiana, the NWD azimuth relative to true north is about 36°, while the SWD azimuth is approximately 144°. Of the three 1064 nm lasers employed by GEDI, one is split into two half-power beams (coverage beams), while the other two remain at full power (power beams). The beams are then slightly dithered, and this setup generates eight parallel ground tracks along the ISS's orbital path: four from the coverage beams and four from the power beams. For a given pass, whether NWD or SWD, the eight parallel beam ground transects cover a swath of 4.2 km, with footprint samples spaced approximately every 60 m along-track and beam transects spaced approximately 600 m apart on the Earth's surface in the cross-track direction (Figure 1b).

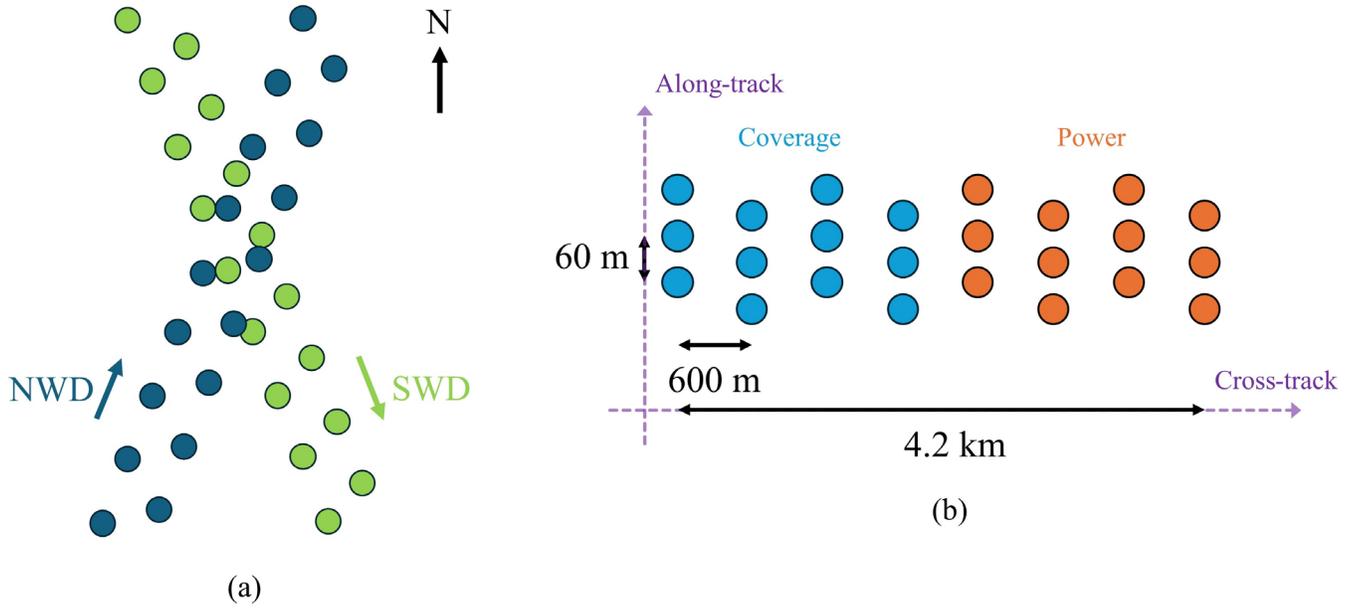

**Figure 1.** (a) GEDI acquisition track azimuthal directions NWD and SWD. (b) GEDI ground sampling pattern.

The GEDI L2A product provides elevation and height metrics derived from the GEDI L1B product's geolocated and smoothed waveforms [47]. These metrics are obtained from six different signal processing configurations or algorithm setting group, with group number 5 generally offering the highest accuracy in tropical areas due to its lower waveform signal end threshold, which better distinguishes weak ground returns in the waveforms [41]. Given the dense vegetation in French Guiana, the *rh_95* metric extracted under the conditions of algorithm setting group number 5 was therefore selected as the direct reliable proxy for canopy height. Between April 2019 and May 2022, 11,798,179 GEDI shots over French Guiana were collected from NASA's GEDI Level 2A Geolocated Elevation and Height Metrics product. These data were filtered to remove irrelevant or erroneous entries, including shots with no detected modes, pure noise signals, and incomplete waveforms [33]. The detailed filters used in this study to remove low-quality observations are described in [41]. Additionally, only GEDI shots with a signal-to-noise ratio (SNR) greater than 10 dB were retained [48]. After filtering, 2,127,076 waveforms remained, which amounts to about 18% of the original dataset. The corresponding *rh_95* values were rasterized on a 10-m grid. For each GEDI shot, the associated *rh_95* value was assigned to a unique pixel corresponding to the center of the footprint. All spatial datasets used in this study were projected to the Universal Transverse Mercator (UTM) coordinate system, Zone 22 North (EPSG:32622), which is the standard projection zone for French Guiana. This UTM projection was selected to ensure accurate representation of spatial relationships and distances in this region. GEDI-derived canopy heights (*rh_95* values) are expressed relative to the WGS84 ellipsoid, as specified by the GEDI L2A product specifications.

*2) Satellite Remote Sensing Input Data*

The canopy height prediction frameworks analyzed in this study leverage both optical and radar remote sensing data to derive canopy height estimates at a spatial resolution of 10 m. Specifically, optical data from the Sentinel-2 (S2) satellite and radar data from Sentinel-1 (S1) and ALOS-2 PALSAR-2 were utilized. The S2 data provided spectral information crucial for understanding vegetation characteristics, through the inclusion of ten spectral bands: B2 (Blue), B3 (Green), B4 (Red), B5-B6-B7 (Vegetation Red Edge), B8 (Near Infrared, NIR), B8A (Narrow NIR), and B11-B12 (Short Wave Infrared). A relative reflectance normalization was conducted for each spectral band to harmonize reflectance values between orbits [49]. To ensure consistency with the final 10-m canopy height prediction grid, the S2 bands originally at a 20-m resolution (B5-B6-B7, B8A, B11-B12) were resampled to 10 m using nearest neighbor value assignment. This method preserves the original pixel values without introducing artificial gradients. It was used solely to harmonize and align the input layers spatially, without increasing the information content. Meanwhile, the radar data from S1 (C-band) and ALOS-2 (L-band) offered complementary structural information on the canopy. The S1 data included four layers based on polarization and orbit: VV ascending, VH ascending, VV descending, and VH descending, each with a pixel size of 10 m. ALOS-2 PALSAR-2 provided dual polarization observations (HH and HV), initially at a 25-m resolution, resampled to 10 m. In total, the integration of these datasets resulted in 16 input layers. No variable selection or dimensionality reduction was performed prior to modeling. Both Random Forest and U-Net are robust to redundancy in input features, and all layers were retained to preserve the full spatial and contextual information available. This approach is consistent with our previous work [12] and aligns with findings from Lang et al. [9] that emphasized the benefits of preserving raw spectral and spatial signals over pre-filtered features in deep learning-based canopy height models.

**Table I.** Overview of input variables used in the canopy height prediction models.

| Data Source | Type | Variables | Original Resolution (m) | Resampled Resolution (m) |
|---|---|---|---|---|
| Sentinel-2 | Optical | B2, B3, B4, B8 | 10 | 10 |
| Sentinel-2 | Optical | B5-B6-B7, B8A, B11-B12 | 20 | 10 |
| Sentinel-1 | Radar (C-band) | VV asc, VH asc, VV desc, VH desc | 10 | 10 |
| ALOS-2 PALSAR-2 | Radar (L-band) | HH, HV | 25 | 10 |
| Global 30-m HAND | Terrain | HAND | 30 | 10 |
| ONF | Landscapes | FLT | 30 | 10 |

### 3) Environmental Input Data

Additionally, two environmental descriptors were integrated into the models' inputs: the height above nearest drainage (HAND) and the forest landscape types (FLT), both originally available at a 30-m resolution and resampled to 10 m using nearest neighbor assignment. HAND is a normalized digital elevation model which normalizes topography to the relative heights along the drainage network [50]. FLT provide information on the broader ecological context of French Guiana by describing 20 different forest classes [51]. These descriptors provide valuable contextual information that classical remote sensing data cannot fully reveal, thus enhancing the canopy height estimates by incorporating aspects related to hydrological and geomorphological characteristics of the landscape that directly impact forest dynamics [12].

### 4) ALS Test Data

ALS data were utilized as reference ground truth to assess the accuracy of the canopy height maps both before and after RK. These data were collected by the French National Forests Office (ONF) during several surveys conducted across various study sites in French Guiana between 2016 and 2018. The ALS acquisitions had technical specifications that included an average point density of 10-12 points per m², with each laser pulse covering an approximate diameter of 20 cm and a scan angle of ± 30°. From the raw tridimensional point clouds, canopy height models (CHMs) were generated by ONF at a 1-m resolution (maximum height per 1-m grid cell). To align with the resolution of our canopy height maps, these CHMs were aggregated to a 10-m resolution using the maximum value per 10-m grid cell. The maximum value was chosen because it better represents the top-of-canopy signal captured in GEDI waveforms [34], [52], [53].

The ALS acquisitions were conducted over seven distinct areas of interest (Figure 2): Paul Isnard – Crique Serpent (PAUL_CSE), Paul Isnard – Crique Mousse (PAUL_CMO), Paul Isnard – Voltaire (PAUL_VOL), Paul Isnard – Est (PAUL_EST), Counamama (COUN), Coralie (CORA), and Régina (REG). In this study, localized RK was performed separately on each of the seven individual areas highlighted in Figure 2a, and the results were subsequently aggregated to compute overall accuracy and error metrics.

### C. Predicted Canopy Height Maps

#### 1) U-Net Canopy Height Map

The first canopy height map of all French Guiana utilized in this study was generated using a U-Net deep learning model known as *CHNET* [12]. The model was developed to produce a 10-m canopy height map from multi-source remote sensing data as well as ancillary environmental parameters. Overall, the *CHNET* model's inputs consist of 18 layers, incorporating both remote sensing (S2, S1, and ALOS-2) and environmental (HAND and FLT) data to produce canopy height estimates. The reference data for model training and validation consisted of the rasterized GEDI $rh\_95$ metric values.

The U-Net architecture, known for its encoder-decoder structure, was chosen for its ability to leverage multimodal data and to capture both the local and global features of the input data [28], [54], [55], [56]. The encoder part of the network consists of multiple convolutional layers followed by max pooling, which progressively reduce the spatial dimensions while increasing the depth of the feature maps. The decoder part then upsamples these feature maps, combining them with corresponding feature maps from the encoder through skip connections, which help retain spatial information and improve the localization of features. The *CHNET* model was trained using a sequential scenario approach fully described in [12]: (1) Initial training with raw GEDI data, which involved training the model using the original $rh\_95$ dataset as reference heights; (2) Enhanced GEDI data integration, which involved using a refined GEDI dataset to remove data points deemed likely to be unreliable; (3) Incorporation of environmental descriptors, which added hydrological and geomorphological descriptors HAND and FLT; (4) Geolocation correction of GEDI footprints, which incorporated an iterative geo-correction process to address spatial inaccuracies in the geolocations of GEDI waveforms. Each scenario brought improvements in model accuracy and this process finally resulted in the most accurate canopy height predictions with reduced error and minimal bias.

#### 2) Random Forest Canopy Height Map

In addition to the *CHNET* model, this study also exploits an alternative canopy height map that was obtained from a Random Forest algorithm, hereby referred to as *RFH* (Random Forest for canopy Height estimation). The primary purpose of this additional map is to serve as benchmark for comparison with the *CHNET* model. By including a classical machine learning approach to canopy height mapping, we aim to provide a reference point to better evaluate the performances and advantages of the *CHNET* framework. Random Forest is a widely used ensemble learning method that builds multiple decision trees during training and outputs the average prediction for the individual trees [57]. This method is particularly robust to overfitting while being able to model non-linear relationships for large datasets with many features, making it suitable for our canopy height estimation task.

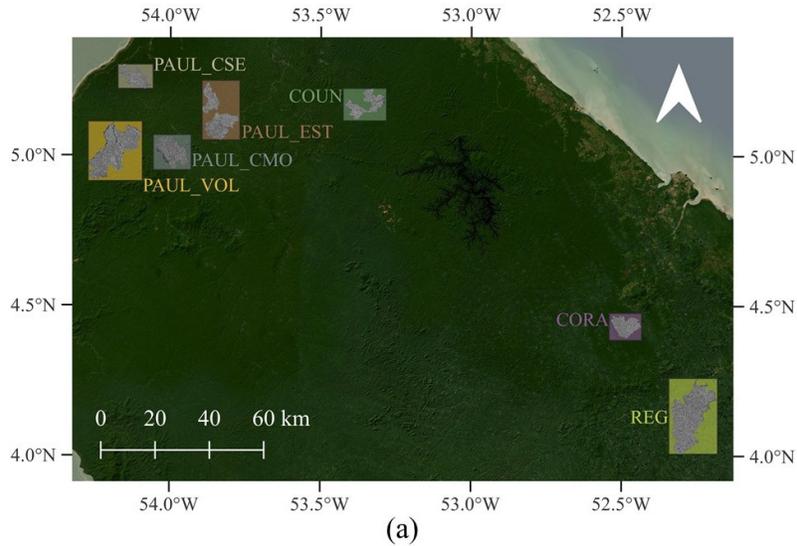

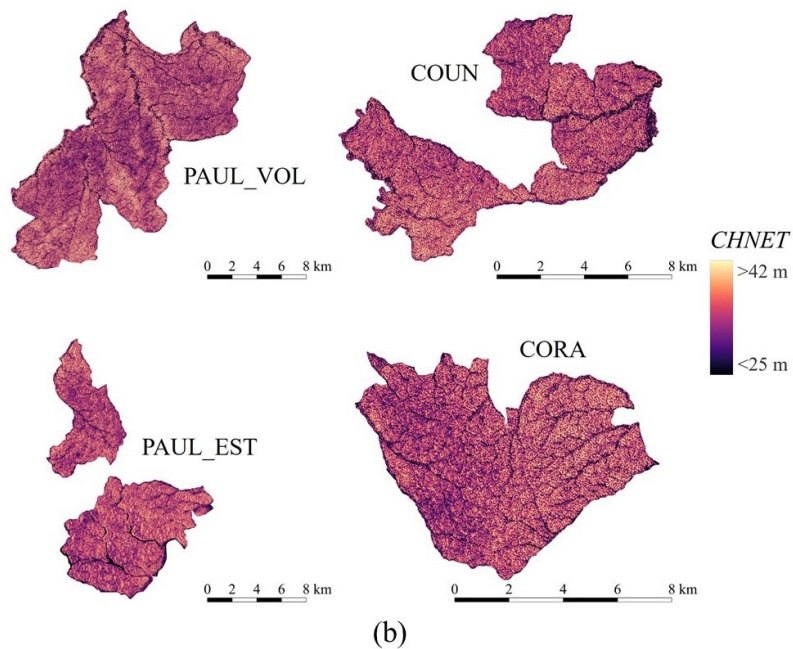

**Figure 2.** (a) ALS acquisition areas of the seven French Guiana study sites (ESRI Satellite®). (b) Detailed canopy height maps (*CHNET* model) at 10-m resolution for four representative study areas across French Guiana: PAUL_VOL, PAUL_EST, COUN, and CORA. These sites illustrate spatial heterogeneity in forest structure across different regions.

For consistency, *RFH* was implemented and tested under the same conditions as *CHNET*, using the same input data and the same geo-corrected GEDI reference dataset. This input data comprised 18 layers (see Table I). Each pixel where a valid *rh_95* value was available was treated as an individual object described by these 18 variables. Therefore, the dataset used to train and validate *RFH* is a table of points (i.e., pixels), where each row represents an individual pixel with its associated GEDI *rh_95* reference canopy height, and the columns represent the 18 descriptive variables. To ensure consistency in our experimental setup, we maintained the same train-validation-test split that was used in the *CHNET* model. This approach ensures that any differences in model performance can be attributed to the algorithms themselves rather than differences in the data used for training and validation. A grid search with cross-validation was also performed to optimize the hyperparameters of *RFH*. After the training process, *RFH* was applied to predict the canopy height for each 10-m pixel across the whole French Guiana. This procedure involved using the trained and validated

*RFH* model to output a predicted canopy height value at each geospatial location based on the 18 input descriptors.

*D. Residual Kriging*

  *1) Ordinary Kriging of Model Residuals*

When mapping a continuous variable such as canopy height across a large study site, understanding the spatial behavior of the reference data is crucial for better interpretation of the final predictions. Since *RFH* treats each prediction independently, any spatial correlation in the results can only be due to existing spatial correlations in the input variables. Nonetheless, some of the unexplained variance in *RFH* predictions may be attributable to spatial correlations within GEDI canopy heights. Therefore, a spatial prediction model is insightful when dealing with spatially dependent data. In contrast, the U-Net architecture utilized in the *CHNET* framework leverages local and global features of the input predictors, offering some capability to better capture spatial dependencies. Unlike *RFH*, which treats each prediction independently, the *CHNET* model's convolutional layers can implicitly identify and learn spatial structures from the input data, which enhances the model's performance in a spatially dependent regression task. However, despite these advantages, *CHNET* does not explicitly model the spatial correlation of GEDI reference canopy heights. Consequently, there is also a need to assess if *CHNET* residuals are spatially structured, which would offer the possibility to explicitly model the spatial correlation of these residuals.

Kriging is a group of geostatistical techniques designed to predict the optimal estimate of a spatially distributed variable $V$ at any unsampled location. This estimate is computed as a weighted average of observed values at surrounding points, with the weights being determined by the spatial covariance of the observations. Various kriging methods exist, and the most relevant for this study is RK [58]. RK posits that the target variable $V$ measured at a location $y$ can be represented as the sum of a deterministic trend $d$ and stochastic residuals $s$:

$$V(y) = d(y) + s(y) \quad (1)$$

The deterministic drift $d$ represents the large-scale trend or systematic part of the spatial variable. It is a predictable function that can be modeled using known covariates or polynomial functions. Essentially, it captures the overall structure of the target variable, and it only captures the general trend of the spatial data. It does not account for the more detailed and localized spatial relationships. Conversely, the stochastic residuals $s$ correspond to more detailed local-scale variations that cannot be captured by the deterministic drift. They represent spatially correlated deviations from the large-scale trend, and they contain information that reflects local variability and spatial autocorrelation. In this study, the residuals, defined as the difference between the observed (i.e., the GEDI *rh_95* metric values) and the predicted (i.e., the outputs of either *CHNET* or *RFH*) canopy heights, are estimated by means of an ordinary kriging (OK) procedure. OK assumes second-order stationarity to accurately estimate the spatial covariance structure of the residuals. Second-order stationarity implies that the expectation and the variance of the residuals, considered as a spatial function rather than just as observed data points (i.e., the $s$ function), are constant over space. This means that the expectation and variance do not depend on the specific location $y$ inside the study site. Furthermore, second-order stationarity also implies that the spatial covariance only depends on the separation distance rather than absolute locations. Overall, these assumptions are reasonable for the residuals in the context of this study, as the large-scale trends in canopy heights were removed through model-based predictions (either *CHNET* or *RFH*). Any systematic variation in canopy height due to environmental factors is indeed already captured by the models, leaving the residuals to represent smaller-scale spatial variability that is less influenced by environmental heterogeneity. Although perfect second-order stationarity may not hold in natural ecosystems, and deviations from this assumption can significantly impact the accuracy of kriging estimates [59], it nonetheless serves as a practical foundation for OK and has been commonly employed in forest geostatistical analyses [39], [60], [61], [62].

OK relies on the spatial autocorrelation between nearby values, meaning that nearby locations are likely to have similar values. The estimated residual $\hat{s}$ at an unsampled location $y_0$ is obtained using a weighted average of $n$ known residual values $s$ at nearby sampled locations $y_i$:

$$\hat{s}(y_0) = \sum_{i=1}^{n} \lambda_i s(y_i) \quad (2)$$

The weights $\lambda_i$ are calculated such that the estimation is unbiased, and they are determined by minimizing the estimation variance, under the unbiasedness condition. The values of the weights depend on the spatial structure of the residuals, which is described by means of a semivariogram function [63]. The semivariogram function $\gamma$ quantifies the spatial dependence of the residuals $s$ by measuring the average dissimilarity between pairs of points as a function of the distance $h$ (lag) between them:

$$\gamma(h) = \frac{1}{2N(h)} \sum_{i=1}^{N(h)} [s(y_i + h) - s(y_i)]^2 \quad (3)$$

$\gamma(h)$ represents the semivariance at a lag distance of $h$, while $N(h)$ is the number of pairs of points separated by a distance of $h$ and $s$ are the canopy height prediction model residuals at locations $y_i$ and $y_i + h$.

Empirical semivariograms are a fundamental tool in geostatistics used to characterize the degree of spatial dependence between observations as a function of distance. As highlighted in (3), the empirical semivariogram of the residuals is computed using pairs of points within the study area. Specifically, for each pair of points, the semivariance is calculated as half the average squared difference between the values at the two points. The data is then aggregated into distance bins to average the semivariance values within each bin to smooth out variability and reveal more meaningful patterns. In our study, the width of distance intervals into which data point pairs are grouped for semivariance estimates is set to 100 m. Next, a mathematical function is fitted to the empirical semivariogram to model the spatial structure of the residuals. This fitted function allows the empirical semivariogram of the residuals to be represented mathematically, providing a basis for spatial application in any coordinates of the area. The fitted semivariogram has three key parameters: (1) Nugget, which represents the semivariance at a lag distance approaching zero, indicating measurement error or small-scale variability; (2) Sill, which is the semivariance value where the semivariogram reaches a plateau, beyond which there is no longer spatial correlation; (3) Range, which is the distance at which the semivariance reaches the sill, marking the extent of spatial correlation.

Using the fitted semivariogram, an OK procedure is performed on a 10-m grid to predict residuals at unsampled locations, producing an interpolated map of the model residuals, referred to as kriged residuals. The final step of RK consists in combining the regression predictions $\hat{d}$ with the kriged residuals $\hat{s}$ to produce new canopy height estimates $\hat{V}$ at every unsampled location $y_0$:

$$\hat{V}(y_0) = \hat{d}(y_0) + \hat{s}(y_0) \tag{4}$$

This whole process, comprising the computation of the empirical semivariogram, the fitting of a mathematical function, and the OK procedure, is applied to both *CHNET* and *RFH* residuals. By integrating the kriged residuals back into the initial canopy height predictions, we aim to account for the spatial dependencies not captured by the original models, and thus incorporate both the deterministic and spatial components of the data. *RFH* can only capture the spatial autocorrelation present in the input variables, but not that of GEDI data. Therefore, we use RK to account for the spatial autocorrelation of canopy heights, resulting in an enhanced model referred to as *RFH-RK*. For *CHNET*, although it inherently integrates spatial information through its architecture, we apply RK to evaluate whether additional spatial modeling offers any further benefits, resulting in a new model called *CHNET-RK*. RK allows us to produce new and potentially enhanced canopy height maps on one hand, and to determine the effectiveness of adding a spatial component to both models on the other hand. Ultimately, our analysis includes a comprehensive comparison of different canopy height maps produced with different methodological approaches and subsequently refined with RK.

*2) Spatial Analysis of Model Residuals*

To explore the spatial dependencies of GEDI measurements and understand whether they are accounted for in the regression models, we perform a spatial analysis of GEDI *rh_95* values, regression results, and models' residuals. The primary focus is on the residuals, with the goal of producing for each model a corrected map that incorporates the spatial correlations of GEDI measurements that were not captured by the regression models. The analysis of GEDI *rh_95* spatial correlations helps confirm that the observed correlations in the residuals are indeed associated with the GEDI data themselves, rather than being a result of the regression models. Similarly, the predictions from *CHNET* and *RFH* are also analyzed. These spatial analyses are based on the computation of empirical semivariograms to describe spatial patterns. These empirical semivariograms provide valuable information about the spatial structure and variability of the data, revealing patterns such as spatial continuity and potential anisotropies.

The specific parameters of the GEDI acquisition, i.e., sparse measurements along azimuthal tracks and laser beam type, introduce a spatial correlation in the regression results that does not reflect the actual canopy structure and dynamics. We hypothesize that the acquisition configuration as well as the characteristics of the GEDI sensor introduce anisotropies originating from the measurements themselves rather than the canopy. To perform effective RK based on a representative spatial structure, it is crucial to differentiate and account for the spatial correlations of forest canopy heights, while minimizing or correcting for the spatial correlations introduced by the instrument or measurement methodology.

Given the specificities of the GEDI data used as reference canopy height and their potential spatial effects we aim to observe, we first compute empirical semivariograms at different levels of GEDI data filtering with respect to beam type, including all data, power-only, and coverage-only. This approach isolates each particularity of the GEDI acquisition in terms of laser energy to understand its impact on the spatial autocorrelation of the measurements. Laser beam energy can be a significant source of anisotropy and inconsistency in the measurements, as laser energy is halved depending on beam type. Additionally, directional semivariograms are computed to study the spatial correlation in specific directions. We hypothesize that the azimuthal configurations of GEDI tracks introduce anisotropic effects, particularly in the cross-track directions. Indeed, each ground track corresponds to either power or coverage beams. Moreover, they are acquired at different periods, and the atmospheric conditions at the time of the acquisitions may differ, resulting in different received waveforms, and in turn eventual changes in the *rh_95* values across tracks. To validate this hypothesis, we perform a spatial analysis through empirical semivariograms computed from different levels of GEDI data filtering with respect to track azimuthal direction. Specifically, semivariograms are computed using data from NWD tracks on one hand, and SWD tracks on the other, which are subdatasets containing parallel tracks only. We hypothesize that the true forest canopy heights should exhibit minimal anisotropy, meaning that there should be no significant change in the spatial correlation of canopy height with azimuthal direction when averaged over a large area such as French Guiana. Consequently, any observed anisotropy in the residuals (or in *rh_95*) is likely attributable to sensor-related effects rather than natural properties of the forest. Eventually, we consider that the spatial structure of the residuals characterized under conditions that mitigate anisotropic effects reflects the reality of the horizontal structuring of canopy height in all directions in an isotropic manner.

In the end, these spatial analyses are a crucial step to obtain a relevant and reliable representation of the spatial structure of the residuals, with the final goal of performing RK. In this study, using the insights gained from these preliminary analyses, the empirical semivariograms of the models' residuals are computed under conditions designed to mitigate GEDI sensor-induced anisotropies and to represent the true spatial correlations of forest canopy heights.

*3) RK Procedure*

The ultimate goal of the analyses previously presented is to guide the construction of reliable empirical semivariograms for the kriging process. From the insights gained from these analyses, we first retain only GEDI power beam measurements for the computation of empirical semivariograms, which serve as basis for the subsequent kriging procedure. This allows mitigating inconsistencies related to laser energy. Second, to address anisotropies arising from orbital acquisition geometry, we separate GEDI power beam data by track azimuthal direction (NWD and SWD subdatasets) and compute directional semivariograms for each. Directional semivariograms are an effective tool to study spatial patterns in given directions. Specifically, semivariograms are calculated in the along-track and cross-track directions for each subset. Only the along-track semivariograms from a single azimuthal configuration (NWD or SWD) are used in the kriging process, as they provide a more reliable empirical basis for modelling spatial dependencies. To reduce directional bias in the semivariance estimates, we use a strict tolerance angle of 1° when computing directional semivariograms. Although this constraint reduces the number of point pairs used per distance bin and may introduce noise, it ensures that spatial correlations are characterized along precise directions to isolate meaningful patterns.

These methodological choices are made to produce relevant empirical semivariograms of the models' residuals that accurately represent the actual spatial correlation of GEDI canopy heights in French Guiana. The next step in the RK procedure is to fit mathematical functions to these empirical representations. In this purpose, many classical functions exist, such as linear, spherical, circular, gaussian, and exponential models, each characterizing different types of spatial correlation. A linear model means that the spatial autocorrelation increases linearly with distance. The spherical model is a modified quadratic equation where the spatial dependence levels off at the sill and range values. Circular models resemble the spherical ones, with spatial dependence diminishing to an asymptotic level. The gaussian models use a normal probability distribution curve and have an inflection point. For exponential models, the spatial autocorrelation gradually reaches the sill, with the relationship between two distance bins decaying progressively until spatial dependence dissipates at an infinite distance.

The fitted semivariograms serve as the covariance models used to perform RK for *CHNET* and *RFH*. The goal of the RK procedure is to improve canopy height predictions by incorporating spatial autocorrelation not captured by the regression models. The method consists in interpolating the residuals from the *CHNET* and *RFH* models using OK and adding the resulting kriged residuals to the original model predictions. To explore the practical application of RK and its effectiveness in correcting the regression models' predictions, we first conduct localized computations rather than applying the method across an entire study site. The goal of this preliminary step is to have an initial assessment of the capabilities and potential of the RK procedure in smaller zones before scaling up. By initially focusing on localized areas, we are able to gain insights into the kriging process and draw preliminary observations. Kriging is then applied separately to each of the seven study areas where ALS data are available (see Figure 2a). These areas are treated independently because they are spatially separated by distances larger than the range of spatial autocorrelation identified in the semivariograms. For each study site, a 10-m resolution grid is used for kriging to be consistent with the resolution of the canopy height maps. Prior to kriging, we introduce an additional buffer zone of 3 km around each study site to extend the areas on which kriging is performed. This buffer serves to mitigate edge effects when comparing the predictions corrected through kriging with ALS canopy heights. Indeed, spatial interpolation near the edges can be less reliable due to the lack of nearby sampled points. Interpolating on buffered areas ensures that values on the edges do not compromise the comparison with the test data. The 3-km value corresponds approximately to the range of spatial autocorrelation observed in this study.

For kriging, we only use GEDI residuals from power beams located within the buffered study areas. Residuals are computed as the difference between GEDI *rh_95* reference values and the predicted canopy height (from either *CHNET* or *RFH*). OK is performed using the fitted semivariogram function associated with each regression model, and the output is a 10-m continuous map of kriged residuals over the full buffered grid. These kriged residuals are added to the original regression predictions at each pixel of the grid, thus producing spatially corrected canopy height maps. The buffered margins are then removed to produce final maps aligned with the original ALS study zones. This workflow is executed consistently for each study site and for both regression models, using the same parameters and semivariogram models across all cases.

*4) Validation Strategy*

A visual summary of the complete methodological workflow is presented in Figure 3. The accuracy of the predicted canopy height maps before and after kriging is evaluated against both ALS data and GEDI measurements. The assessment is based on two standard performance metrics: RMSE and mean bias, computed at the pixel-level. The relative RMSE (rRMSE), which is the RMSE normalized with the mean of the observed values, is also used. Validation is performed by aggregating all seven study sites into a single dataset. This approach is possible because the distance between the different areas of interest is greater than the range of the fitted semivariograms, which ensures that each site can be treated as an independent kriging domain. As a result, the interpolation performed within each site does not influence others, and the prediction errors from all sites can be combined without introducing spatial dependence. Moreover, the study areas are predominantly composed of tall canopies, with limited representation of low vegetation and trees, resulting in relatively low structural variability both within and across the ALS study sites. This aggregated evaluation strategy has the advantage of increasing the number of validation samples, which allows improving the statistical robustness of the performance metrics. While detailed site-level analyses are valuable for understanding local spatial behaviors (see Results), aggregating the performance metrics offers a more concise and interpretable summary of model accuracies.

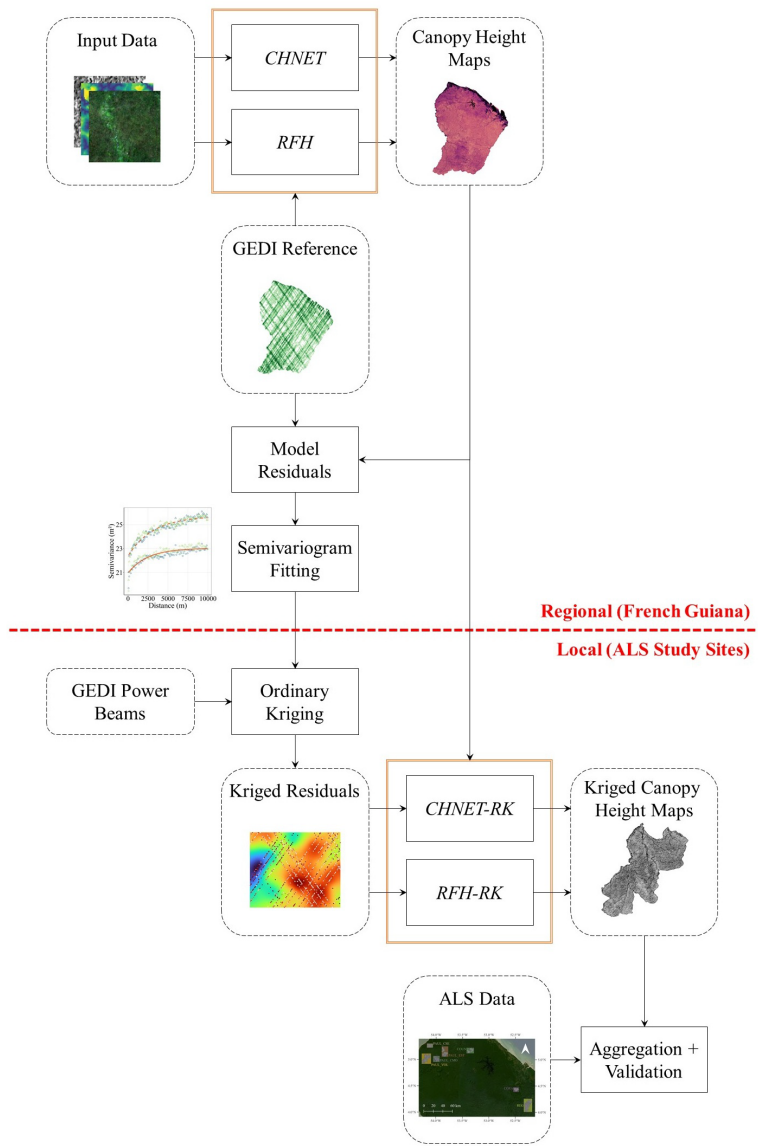

**Figure 3.** Overview of the methodological workflow. *CHNET* and *RFH* models were trained using multisource input data and GEDI reference canopy heights to produce initial canopy height maps. Model residuals were computed, and directional semivariograms were fitted. OK was then applied to the residuals using GEDI power beams within local ALS study areas. Kriged residuals were added to model outputs to generate corrected canopy height maps (*CHNET-RK* and *RFH-RK*), which were subsequently validated against ALS data.

Additionally, beyond standard accuracy assessment, our validation strategy also aims to explore the practical conditions under which RK can bring benefit to regression models. When assessing the results of *CHNET-RK* and *RFH-RK* against GEDI data, the evaluation is performed at discrete points, whereas assessing the results against ALS data involves the use of continuous maps. Consequently, any improvements introduced by RK might be less apparent in an overall comparison with ALS data, as the enhancements could be less discernible when averaged across the broader spatial extent of the ALS acquisitions. In particular, we examine how the proximity and density of GEDI observations affect kriging effectiveness. We conduct a proximity-based analysis around GEDI power beam shots used in the kriging process. Specifically, we define concentric circular buffers of increasing radius centered on each GEDI footprint and compute errors for model predictions (of *CHNET-RK* and *RFH-RK*) within each buffer. By progressively increasing the radius, we assess how prediction improvement evolves as a function of distance from the nearest GEDI observation. This approach allows us to characterize how kriging performance is affected by the availability and spatial density of GEDI observations. It also helps to quantify the approximate distance within which interpolation through kriging can yield significant improvements. Beyond this threshold, the lack of available nearby observations may indeed limit the effectiveness of kriging in correcting prediction errors.

## III. RESULTS

### A. Semivariogram Analysis

*1) CHNET Predictions and Residuals*

We first computed and plotted the empirical semivariogram of the reference heights used for the *CHNET* model's training and validation, specifically the *rh_95* values (Figure 4a). The semivariogram of the GEDI *rh_95* data exhibits a pronounced spatial periodicity of approximately 600 m. This periodicity is particularly strong at the 600-m mark, where the semivariance peaks significantly. This trend continues, although with a diminishing effect, at subsequent multiples of 600 m. This 600-m periodicity is consistent with the distance between two adjacent GEDI ground tracks, indicating a strong spatial dependence corresponding to the sensor's acquisition pattern.

Similarly, we computed the semivariogram for *CHNET* predictions of canopy height and corresponding model residuals (Figure 4b). The predictions display a clear spatial structure, which means that *CHNET* accounts for some spatial correlation in its predictions. The semivariogram of the residuals also shows a degree of spatial structure, with slightly lower semivariance values than the predictions. This implies that while *CHNET* accounts for some spatial patterns, a portion of the spatial autocorrelation remains unexplained by the model and is contained within the residuals. The shape of the empirical semivariograms of the residuals suggests stationarity, as they stabilize at a sill with no increasing trend over long distances. This observation further strengthens the assumption of stationarity on which the kriging procedure relies. The results also reveal the same periodic pattern for the residuals as the one observed in the semivariogram of GEDI *rh_95* data. Observing the same periodicity in the semivariograms of both GEDI *rh_95* values and the residuals supports the idea that this periodicity is linked to GEDI and is not an artifact from the *CHNET* model. This idea is further confirmed by the semivariogram of the predictions, which indicates that the model performs some level of smoothing on the periodicity present in the semivariogram of GEDI *rh_95* data. Therefore, we conclude that the observed periodicity is sensor-related and is not linked to a specific spatial organization of the forest canopy height. Specifically, the 600-m spatial pattern reflects the characteristics of the GEDI sensor and its data acquisition process rather than the behavior of the *CHNET* model itself.

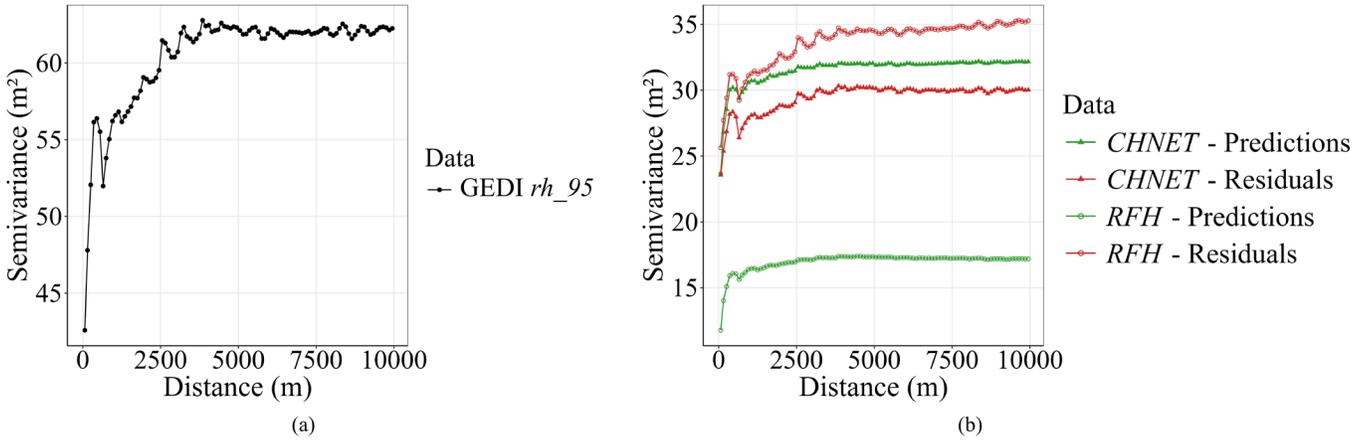

**Figure 4.** (a) Omnidirectional semivariogram of GEDI *rh_95* values. (b) Omnidirectional semivariograms of *CHNET* and *RFH* predictions and residuals.

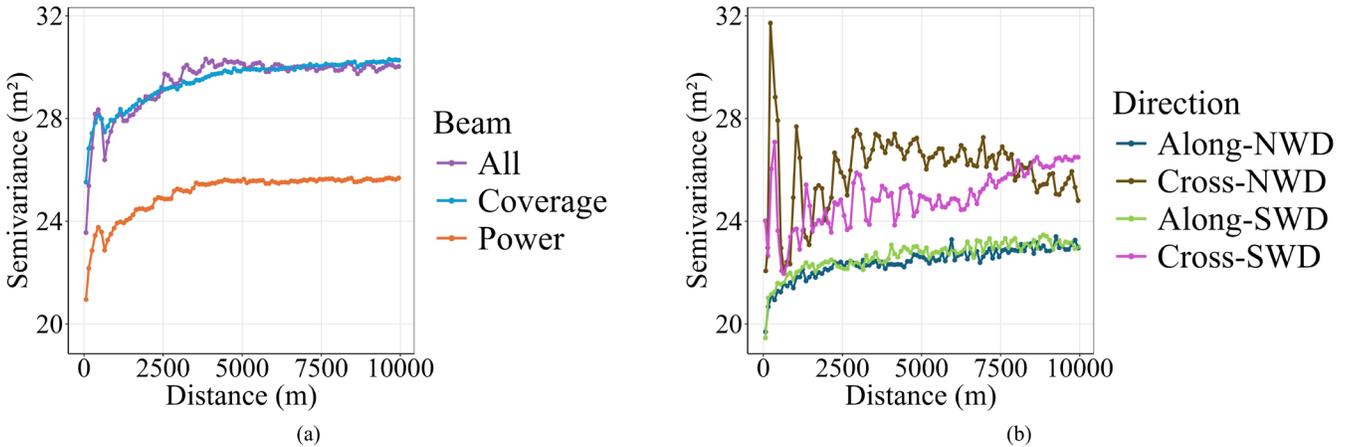

**Figure 5.** (a) Omnidirectional semivariograms of *CHNET* residuals depending on beam type. (b) Directional semivariograms of *CHNET* residuals in the along-track and cross-track directions.

Regarding laser energy, we plotted the empirical semivariograms of *CHNET* residuals based on beam type (Figure 5a). We found that the magnitude of semivariance values differs substantially depending on beam type, with significantly lower values for power beams compared to coverage beams. Additionally, when considering a single beam type in the spatial analysis, the observed periodicity is notably less pronounced than when all beams are included. These findings indicate that differences in laser beam type can be considered as a source of anisotropy in GEDI measurements, and that the associated spatial autocorrelation does not reflect the actual spatial structure of the forest's canopy. Therefore, only GEDI shots of full power (i.e., power beams) are retained for the subsequent steps of the study.

Next, we plotted the directional semivariograms of *CHNET* residuals in the along-track and cross-track directions for each azimuthal configuration (NWD and SWD). Figure 5b shows the directional semivariograms in the along-NWD and cross-NWD directions, which correspond to 36° and 126° relative to true north, respectively, and in the along-SWD and cross-SWD directions, which correspond to 324° and 234° relative to true north, respectively. Notably, the results obtained for NWD and SWD azimuths are equivalent. Cross-track semivariograms exhibit a high periodicity, while the periodicity is minimal in the along-track directions. These directional semivariograms clearly demonstrate that the spatial periodicity previously observed in Figures 4 and 5a is linked to the pattern of GEDI beam ground transects. The 600-m period visible in the cross-track directions corresponds to the spacing of GEDI tracks.

The spatial autocorrelation of the residuals is almost equivalent in the along-track directions of both NWD and SWD subdatasets taken independently. Hence, these along-track semivariograms represent the underlying isotropic (i.e., omnidirectional) spatial autocorrelation of model residuals without sensor-induced anisotropies. These semivariograms obtained along-track accurately reflect the actual spatial autocorrelation structure of the residuals. In the end, they can be used isotropically with GEDI power beams in the RK process. Figure 6 shows the associated directional semivariograms of *CHNET* residuals for each subdataset in the corresponding along-track direction. It is important to note that the directional semivariograms presented in Figure 5b and 6 are noisier due to the strict tolerance angle of 1° used during their computation. However, despite this increase in noise, the spatial correlation remains observable, and the validity of the observed trends is not compromised.

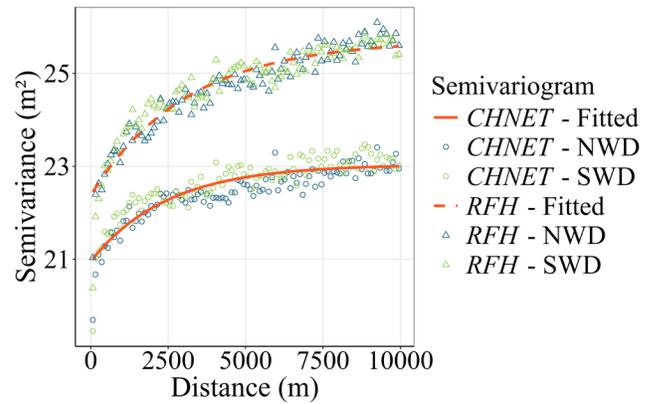

**Figure 6.** Directional semivariograms (empirical and fitted) of *CHNET* power beam residuals for NWD and SWD configurations.

**Table II.** Parameters of the fitted exponential semivariograms for *CHNET* and *RFH* residuals.

| *Model* | *R²* | *Nugget (m²)* | *Sill (m²)* | *Range (m)* |
|---|---|---|---|---|
| *CHNET* | 0.87 | 21.0 | 23.0 | 2466.0 |
| *RFH* | 0.93 | 22.4 | 25.7 | 3096.2 |

*2) RFH Predictions and Residuals*

A comprehensive analysis identical to that performed for the *CHNET* model was also conducted for the *RFH* model. The empirical semivariogram of *RFH* residuals (computed from all available GEDI data) reveals a similar periodic pattern (Figure 4b), which further confirms our earlier findings that the observed anisotropies are attributable to the GEDI sensor itself. Furthermore, the semivariogram of *RFH* predictions (Figure 4b) shows almost no spatial structure. In contrast, the residuals present a significant spatial structure, with significantly higher semivariance values compared to those of the predictions. These residuals integrate the remaining spatial autocorrelation that is not accounted for by the model and are the most spatially structured among all configurations presented in Figure 4b.

Figure 6 presents the directional semivariograms of *RFH* residuals in the corresponding along-track direction for each GEDI power beam subdataset, i.e. either from NWD or SWD tracks. The results corroborate the findings from the *CHNET* residuals analysis. Indeed, to perform relevant RK of the *RFH* model's predictions, the optimal approach also involves using the directional semivariogram of power beam residuals belonging to a single azimuthal direction, which is computed in the corresponding along-track direction.

*B. Residual Kriging*

*1) Semivariogram Fitting*

To perform RK, we fitted mathematical functions to the empirical semivariograms. As highlighted in Subsection III.A,

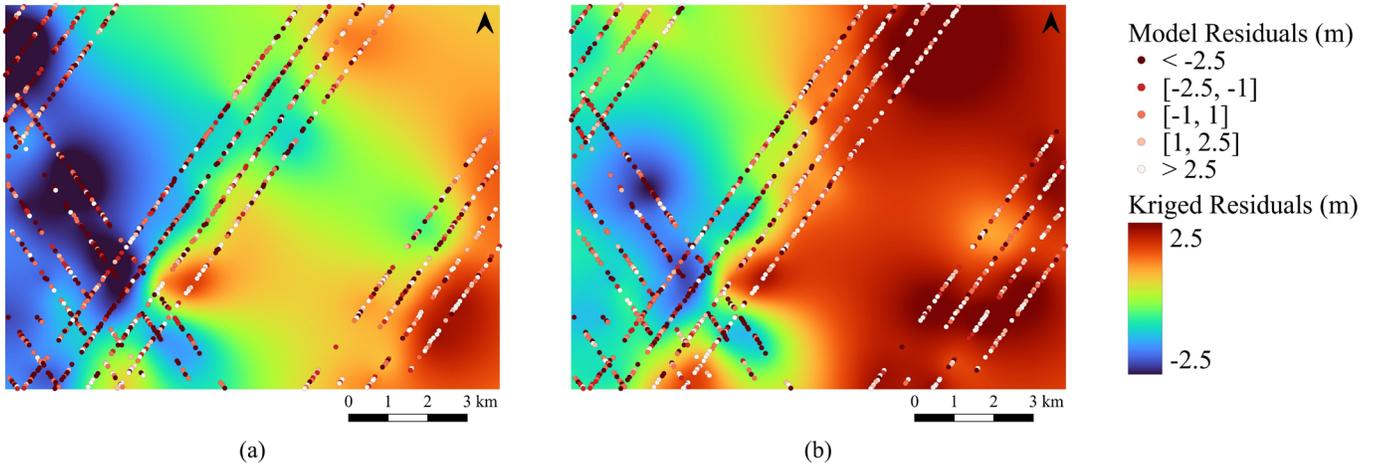

**Figure 7.** Kriged residuals and model residuals of GEDI power beam footprints for *CHNET* (a) and *RFH* (b) in a window within the PAUL_EST area.

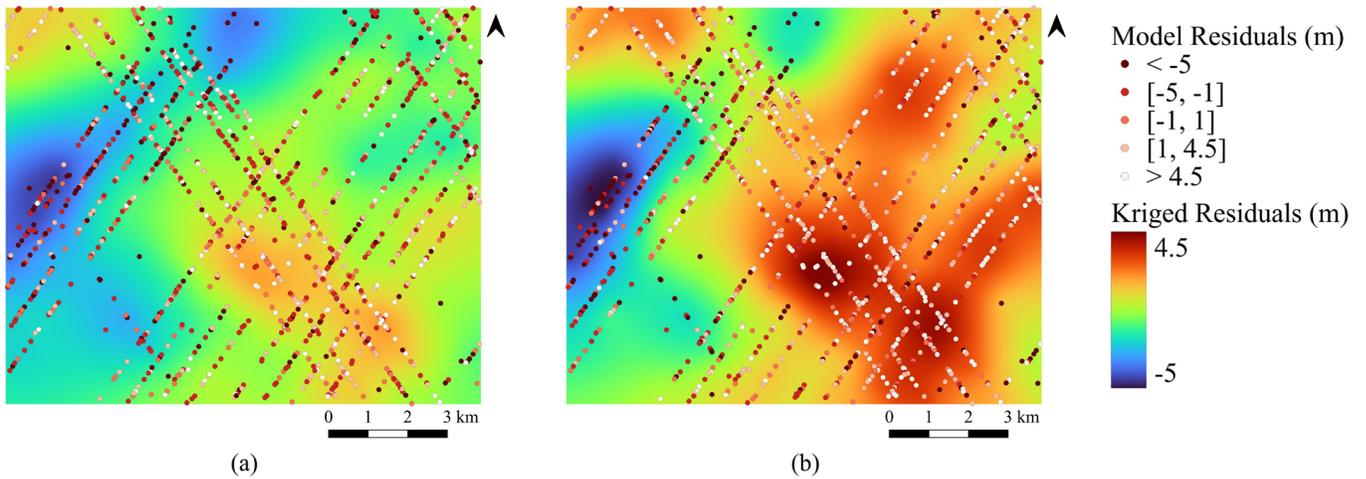

**Figure 8.** Kriged residuals and model residuals of GEDI power beam footprints for *CHNET* (a) and *RFH* (b) in the CORA area.

for each regression model, we generated two empirical semivariograms of the residuals to model the spatial autocorrelation of canopy heights: one for the NWD configuration and another for the SWD configuration. Since both semivariograms displayed equivalent behaviors, which indicates consistent spatial patterns in both configurations, we proceeded by fitting a single semivariogram model to the combined empirical data. We observed that the exponential function exhibited the best correlation with the empirical data, achieving R² scores of 0.87 and 0.93 for *CHNET* and *RFH*, respectively. As shown in Figure 6, the fitting process resulted in exponential models characterized by the parameters described in Table II. The residuals of *RFH* appear to exhibit more spatial structure than those of *CHNET*, as highlighted by the greater sill value. The range of the spatial autocorrelation is also larger by more than 500 m for *RFH* compared to *CHNET*. Spatial dependencies persist over greater distances in the *RFH* residuals, which suggests that the *RFH* model leaves more of the spatial correlation unaccounted for.

*2) Kriged Residuals*

RK was first executed in a localized domain using the fitted exponential semivariograms, and with sampled points consisting of the GEDI power beam footprints. Figure 7 shows a representative example of kriged residuals for *CHNET* (Figure 7a) and *RFH* (Figure 7b) after RK was applied to a localized window of 10×12.5 km in the PAUL_EST area.

Several key observations emerge from this analysis. Firstly, the corrections provided by RK are predominantly localized around GEDI measurement points. These localized corrections appear as patches in the kriged residuals maps (Figure 7), highlighting areas where the kriging process has estimated the residuals based on nearby GEDI data. Secondly, the kriged residuals show a tendency to reproduce the spatial pattern of model residuals. Specifically, in regions where residuals are high and positive, the kriged residuals also tend to be positive. Conversely, in areas where residuals are highly negative, the kriged residuals generally fall within the negative range. This indicates that RK is effectively capturing and adjusting for local variations in the residuals. Thirdly, regions with few or no GEDI footprints tend to have kriged residuals that are closer to the average of the range of values, rather than exhibiting extreme values. This observation underscores that, contrary to regions with many GEDI measurements, areas lacking GEDI information do not experience a significant correction of their predictions. Finally, we note that the magnitude of the corrections applied by RK in this area is similar for *CHNET* and *RFH*, with kriged residuals varying between -2.5 and 2.5 m.

Following this initial assessment, the RK procedure was scaled up and applied across each of the seven ALS study sites. For example, CORA area is characterized by a greater density of GEDI points than PAUL_EST. Figure 8 shows the kriged residuals of *CHNET* (Figure 8a) and *RFH* (Figure 8b) after performing RK. The results allow drawing the same conclusions as before. Specifically, we observe that the magnitude of the corrections applied by RK is less pronounced for *CHNET* than for *RFH*, with kriged residuals varying between -4.5 and 2 m for the former, and between -5 and 4.5 m for the latter. This wider level of correction shows that a more significant adjustment is needed for *RFH* predictions compared to those of *CHNET* in the CORA area. This observation is consistent with the higher model residuals observed for *RFH* (Figure 8), which indicates that *RFH* predictions require a more pronounced spatial correction.

Overall, across all study areas, the kriged residuals ranged from approximately -5.2 m to 4.8 m for *RFH*, and from -4.2 m to 2.6 m for *CHNET*. The magnitude of kriged residuals is generally greater for *RFH* than for *CHNET*, and the corrections remain localized around GEDI footprints.

*3) Enhanced Predictions and Performance Assessment*

Predicted canopy height maps were produced across all study sites for both the original regression models (*CHNET* and *RFH*) and their kriging-enhanced versions (*CHNET-RK* and *RFH-RK*). These outputs were then aggregated for performance assessment. Table III presents a comparison of the four distinct canopy height products: the initial predictions from both *CHNET* and *RFH*, and the spatially corrected predictions obtained from RK for each model (*CHNET-RK* and *RFH-RK*). Figure 9 shows the distribution of differences between model predictions and GEDI before and after RK.

The evaluation of the canopy height maps produced through RK for both *CHNET* and *RFH* reveals several key insights. The primary observation is that RK has a positive impact on the accuracy of canopy height estimates with respect to GEDI data, as it allows reducing the RMSE by half a meter for *CHNET* and about a meter for *RFH*. More specifically, in the case of *RFH*, the RK procedure results in a reduction in bias, as we observe a shift from negatively biased estimates (with a bias of -1.3 m for

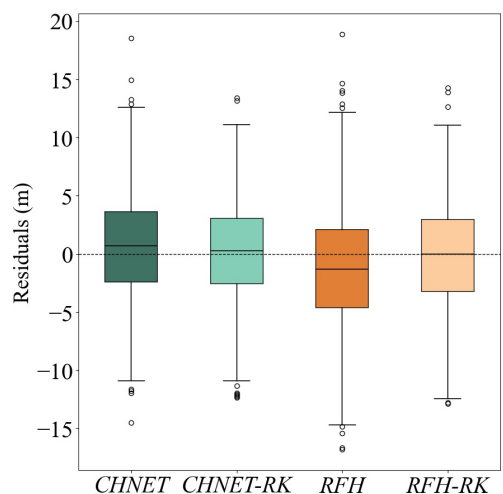

**Figure 9.** Distribution of differences between model predictions and GEDI reference canopy heights before and after RK. Positive values indicate overestimation by the models. The boxplots show the median, the 1st and 3rd quartiles, the 10th and 90th percentiles, as well as outliers.

**Table III.** Accuracy metrics of *CHNET* and *RFH* before and after RK compared to GEDI and ALS test data.

| Test Set | Model | Bias (m) | RMSE (m) | rRMSE |
|---|---|---|---|---|
| GEDI | *CHNET* | 0.6 | 4.7 | 13.0% |
| | *CHNET-RK* | 0.0 | 4.2 | 11.6% |
| | *RFH* | -1.3 | 5.3 | 14.7% |
| | *RFH-RK* | 0.0 | 4.4 | 12.2% |
| ALS | *CHNET* | -0.2 | 5.8 | 16.1% |
| | *CHNET-RK* | -0.1 | 5.8 | 16.1% |
| | *RFH* | -1.8 | 6.1 | 16.9% |
| | *RFH-RK* | 0.5 | 6.1 | 16.9% |

*RFH*) to unbiased estimates (no bias for *RFH-RK*). This indicates that RK tends to shift *RFH* estimates upwards to align them more closely with GEDI measurements. The same improvement is observed regarding *CHNET* predictions, although in a less notable way, with a bias value going from 0.6 m for *CHNET* to a null bias value for *CHNET-RK*. Overall, the integration of kriged residuals through an RK procedure proves to enhance both *CHNET* and *RFH* predictions when assessed against GEDI reference canopy heights.

Conversely, a striking observation is that RK seems to have no discernible impact on the general accuracy of canopy height estimates when compared to ALS data (see Table III and Figure 10). Indeed, *CHNET* and *CHNET-RK* exhibit almost identical performances. In the case of *RFH*, the application of RK still results in an upward shift in the estimates, which is evident in the change from a strong negative bias of -1.8 m for *RFH* to a slightly positive bias of 0.5 m for *RFH-RK*. Despite this adjustment, no improvement in RMSE is observed. To better investigate the impact of RK in relation to ALS data, we computed errors within circular neighborhoods around GEDI points with varying radius values.

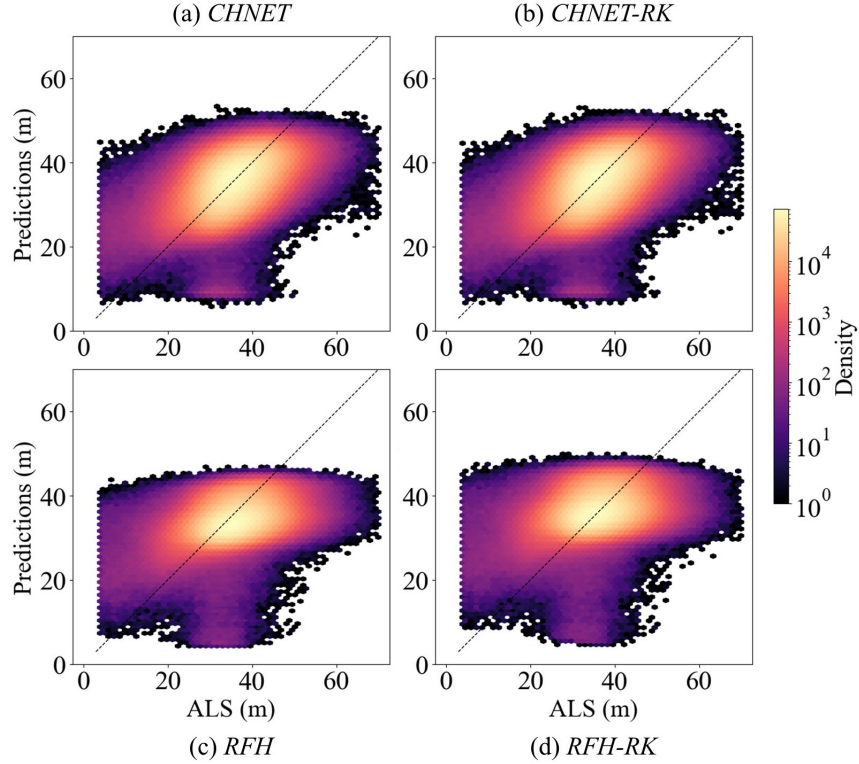

**Figure 10.** Canopy height predictions as a function of ALS ground truth data for *CHNET* (a), *CHNET-RK* (b), *RFH* (c), and *RFH-RK* (d).

Table IV presents the RMSE values of *CHNET-RK* and *RFH-RK* compared to ALS data within circular windows centered around GEDI points. A radius of 0 m represents the exact location of the GEDI points, while an infinite radius corresponds to the full spatial extent of the ALS data. "Range" is a radius value that corresponds to the range of the fitted semivariogram for each model (see Table II), i.e., 2466.0 m for *CHNET* residuals and 3096.2 m for *RFH* residuals. As the radius decreases, both *CHNET-RK* and *RFH-RK* exhibit a progressive decrease in RMSE values, which suggests that the RK process is more effective in regions closer to GEDI points, where more significant corrections may have been applied. The trends are consistent for both *CHNET-RK* and *RFH-RK*, with an overall advantage for *CHNET-RK* across all configurations. Notably, between the point comparison (at a radius of 0 m) and the full ALS extent (infinite radius), there is an approximate RMSE difference of 1 m for both models, which indicates the localized effectiveness of RK at GEDI points. Additionally, at a radius equal to the range of the fitted semivariograms, the RMSEs for both models are the same as the infinite radius. This is due to the spatial density of the GEDI database. Indeed, such radius values around GEDI points almost amount to the extent of the ALS acquisitions, which makes these two configurations identical.

**Table IV.** RMSE values of *CHNET-RK* and *RFH-RK* compared to ALS data within circular windows around GEDI points.

| Radius (m) | RMSE (m) | |
|---|---|---|
| | **CHNET-RK** | **RFH-RK** |
| 0 | 4.9 | 5.0 |
| 250 | 5.2 | 5.4 |
| 500 | 5.3 | 5.6 |
| 1000 | 5.3 | 5.7 |
| Range | 5.8 | 6.1 |
| ∞ | 5.8 | 6.1 |

To illustrate model behavior at a fine spatial scale, we finally present selected canopy height profiles extracted along linear transects in different forest areas (Figure 11). These examples offer a visual comparison between the RK-corrected predictions and ALS ground truth data. Overall, both models follow the general canopy height trends. Random Forest (*RFH-RK*) tends to produce smoother and more averaged predictions, especially over heterogeneous vegetation, compared to U-Net (*CHNET-RK*). Additionally, both models consistently underestimate tall canopies, which is particularly visible in Figure 11a, where predictions do not exceed 40 m even when ALS data clearly indicate higher canopy heights.

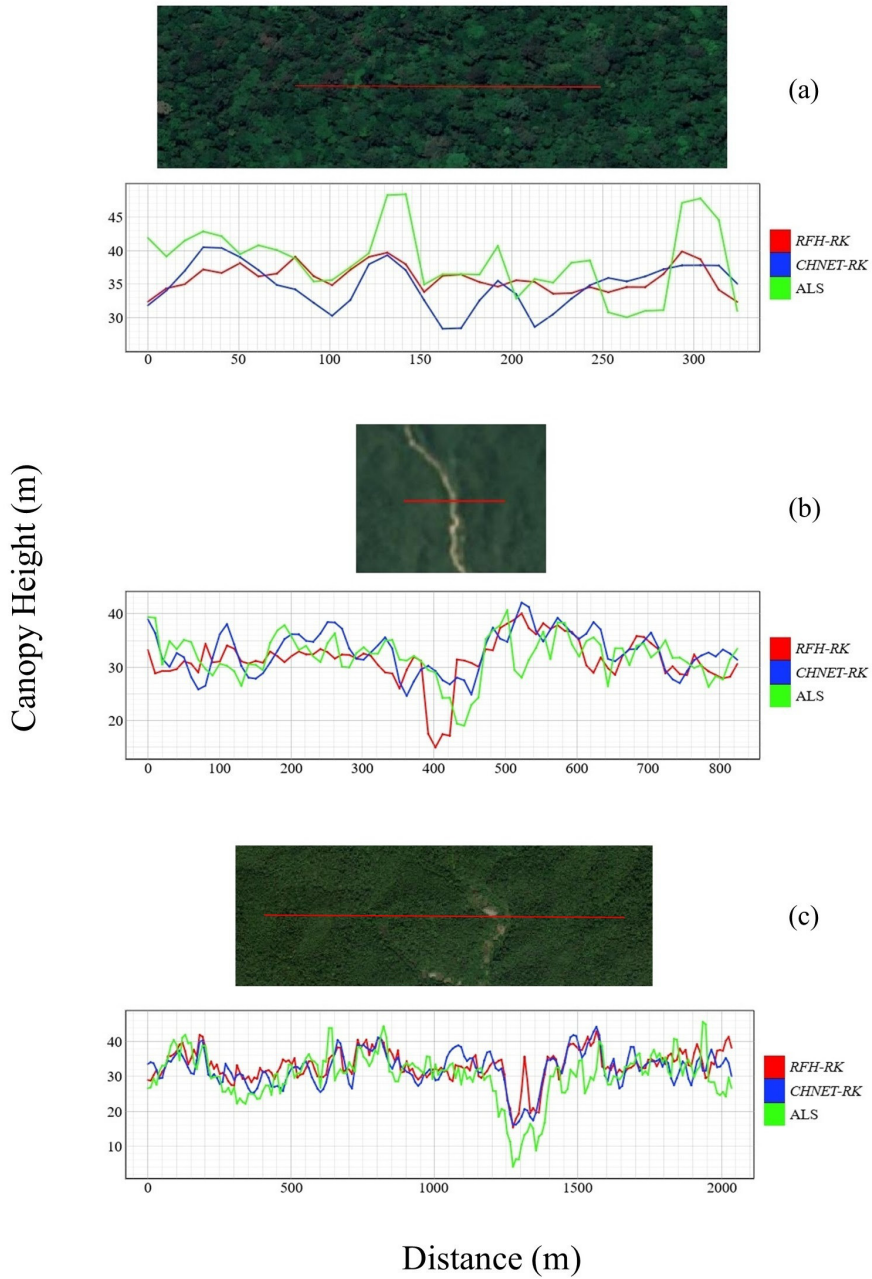

**Figure 11.** Examples of canopy height profiles along selected transects in French Guiana. Each panel shows (top) the transect location (red line) on satellite imagery (Google Satellite® and ESRI Satellite®) and (bottom) the corresponding canopy height profiles comparing ALS ground truth data with *CHNET-RK* and *RFH-RK* predictions. Panel (a) highlights underestimation of tall canopies, while panels (b) and (c) show profiles over more heterogeneous forest structure.

## IV. Discussion

### A. On Sensor-Induced Anisotropies

At the scale of French Guiana, there should be no periodicity related to canopy height in the spatial correlation, nor should there be any anisotropies. The results of our study demonstrated that GEDI's data acquisition parameters introduce sources of spatial anisotropies that are not representative of the actual horizontal structure of the canopy. These anisotropies are linked to the pattern of the GEDI sensor acquisition as well as signal physical parameters such as laser energy. In French Guiana, and more generally in tropical biomes, the penetration of vegetation by the signal is a major challenge, as these ecosystems are characterized by tall and dense canopies. To accurately measure canopy height, the signal must reach the ground, and the capability of the signal to penetrate through the forest and detect the ground is closely related to the laser's physical properties.

Numerous studies have highlighted the strong relationship between signal penetration capabilities and signal properties, which directly impacts the measurements performed by GEDI and the derived canopy heights [52], [64], [65]. Specifically, GEDI's power beams are twice as powerful as coverage beams in terms of laser energy, significantly influencing their ability to penetrate dense canopies. For example, Fayad et al. [32] noted that coverage lasers exhibited substantially lower performances for tree height estimation compared to full power configurations like those of GEDI and NASA's Land Vegetation and Ice Sensor (LVIS). In an analysis over protected study sites in French Guiana, Lahssini et al. [41] observed that power beams produced significantly better measurements for canopy height estimation compared to coverage beams. They found that coverage beams tended to underestimate tree heights, whereas power beams showed a strong linear correlation with ALS reference canopy heights. Coverage lasers generally face more difficulty in reaching the ground because of the dense vegetation, leading to an overestimation of ground elevation and an underestimation of canopy heights. Conversely, power beams can penetrate the vegetation more effectively, and produce return waveforms with a recognizable lowest mode corresponding to the actual ground. Other studies have also confirmed that beam strength significantly impacts the results and have recommended using only power beam data for more accurate canopy height estimates [66]. Nevertheless, GEDI power beams still exhibit a tendency to underestimate canopy height in dense tropical forests with high AGB [32]. Strong beams can also tend to overestimate canopy height in areas of low vegetation. For example, Moudrý et al. [33] found that beams with strong sensitivities (superior to 0.9) typically overestimated canopy height in grasslands. Although a lot of work has focused on the underestimation of tall canopies, accurate observations of low vegetation are equally important for modelling forest structure at regional scales.

These differences in measurements introduce anisotropies because the different beams do not produce the same measurements. For example, the same location in a dense tropical forest observed by a power or coverage laser would produce different waveforms, and consequently, different reference heights for that particular location. When these effects are combined with the ground track pattern of GEDI, which includes straight and parallel beam transects of either power or coverage laser in two azimuthal directions, it creates periodic anisotropy patterns in given directions. Our semivariogram analysis demonstrated that depending on the beams considered and the direction of the analysis, the subsequent spatial autocorrelation extracted was not consistent. Therefore, it is crucial to account for anisotropies that are due to sensor-related effects to analyze the semivariograms of model residuals and eventually perform RK. In this study, the GEDI dataset was gradually refined to achieve a configuration that provided an accurate and artifact-free representation of the spatial correlation of GEDI canopy heights. In tropical contexts, laser power is of paramount importance, which is why we chose to use power beams exclusively to obtain a consistent representation of the spatial structure of canopy heights. We also chose to incorporate a directional aspect in the spatial analysis to account for the spatial sampling of GEDI measurements. An alternative approach to addressing sensor-induced anisotropies would be to handle these issues earlier in the regression process by implementing a preliminary step to refine the GEDI data used as reference for model training. In this approach, the regression model is not responsible alone for managing sensor-related discrepancies, as it is fed with preprocessed and corrected data. For example, although in a different context over temperate forests and using airborne lidar, some studies have underlined the sensitivity of lidar metrics to scan angles and proposed methods for accounting for scan geometry to compute more accurate metrics [67], [68]. By addressing sensor-related effects upfront, the data fed into predictive models become more representative, which leads to more accurate predictions. While this work was conducted in a different biome and with airborne data, the approach of pre-correcting for sensor-induced biases could also be relevant to spaceborne lidar GEDI. Corrected GEDI-derived heights could be obtained by constructing linear or non-linear models based on GEDI metrics, signal parameters, and other environmental factors, like terrain that can also introduce anisotropies in GEDI measurements [34]. Provided there is enough ground truth data, such as ALS data, to build these models, they could be applied to correct GEDI relative height metrics and produce new reference heights that are free from anisotropy biases. In our study, however, we chose to filter GEDI data to retain only the most relevant measurements. Coverage beams present significant challenges, particularly in dense tropical environments, due to their limited penetration of the vegetation. Accurately determining the true values of the associated metrics is difficult, and attempting to correct them would likely introduce additional noise and uncertainty. Furthermore, addressing the 600-m periodicity is even more challenging, as the differences between tracks arise from complex atmospheric factors that are difficult to model and correct in a reliable way. Consequently, it is more effective in our case to filter the data and focus on usable measurements rather than introduce corrections that have their own uncertainties.

*B. On Spatial Information for Canopy Height Mapping*

When mapping a continuous variable like canopy height, it is essential to understand that it exhibits spatial characteristics influenced by numerous environmental and ecological factors. Canopy height is not uniformly distributed across landscapes due to the complex interactions between many different environmental parameters. For example, topography can play a vital role in forest growth by influencing water availability and hydrological networks. In this perspective, the height above nearest drainage (HAND) descriptor proved to provide essential contextual information that is directly related to canopy state, structure, and dynamics. This is further supported by the feature importance analysis of the *RFH* model (Figure 12), where HAND appears as the most important variable by a wide margin. S2 Red Edge 1 and NIR bands also rank highly, which is coherent with their well-known sensitivity to canopy structure. Numerous studies have emphasized the significance of the HAND grid when studying tropical forest ecosystems. For instance, Schietti et al. [69] found a correlation between HAND and changes in floristic composition in the Amazon. Regarding structural parameters, the relationship between HAND and biomass in Eucalyptus plantations in Brazil demonstrated that the functioning and dynamics of the same tree species can vary significantly depending on drainage availability [70]. Therefore, the spatial variability of canopy height is a complex phenomenon, and this spatial dependence needs to be considered in any predictive approach.

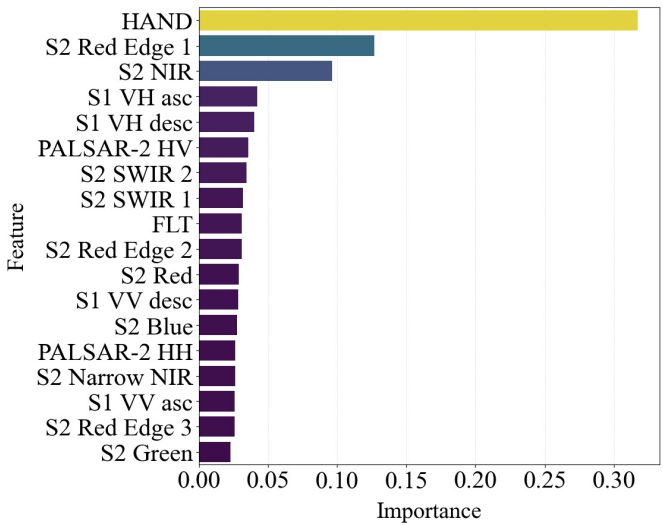

**Figure 12.** Feature importance analysis of the *RFH* model. Importance values are unitless and reflect the relative contribution of each variable to the prediction accuracy (computed from the mean decrease in impurity).

In this perspective, the U-Net architecture on which *CHNET* is built allows accounting for both local and global contexts when mapping canopy height. This architecture enables the model to capture small-scale details and broader spatial patterns in the input data. This ability of U-Net to integrate information from multiple scales through its encoder-decoder structure is particularly beneficial in learning complex spatial relationships in the input data. Consequently, *CHNET* predictions exhibit a spatial structure that is well retrieved by the model (Figure 4b). This explains why the residuals alone do not contain all the spatial autocorrelation, as some of it is successfully captured by the model. However, despite this capacity for spatial feature learning, *CHNET* does not directly incorporate all the spatial autocorrelation of GEDI reference canopy heights. These canopy heights are only used as reference for training and are not fed into the model as part of the input data that *CHNET* uses to make predictions. As a result, the model can learn spatial patterns indirectly but cannot fully represent the spatial autocorrelation of the canopy heights in its predictions. This limitation leaves some unexplained variance linked to the spatial autocorrelation of canopy heights. This remaining spatial variance is indeed found in the spatial structure observed in the residuals (Figure 4b). Implementing an RK procedure can therefore still be relevant and useful for a spatial method like *CHNET*, which is highlighted by the improvements in accuracies presented in this study. Other studies have demonstrated the benefits of combining deep learning frameworks with spatial considerations. Liu et al. [71] employed an interpolation-based mapping strategy, combining multi-layer perceptrons with kriging interpolation, to produce a 30-m canopy height map of China, and they noted that their method allowed reducing the saturation effect of estimates in tall forests.

In their canopy height model of the Earth, Lang et al. [9] used CNNs trained on S2 data and encoded geographical coordinates, and they observed that their model yielded far better results when this geographical information was incorporated compared to S2 data alone. To integrate spatial information, they designed their model with the ability to learn geographical priors, by feeding it geographical coordinates (in a suitable cyclic encoding) as additional input channels. By doing so, the model could better understand spatial relationships and contextual information in relation to S2 reflectance values, leading to more accurate and reliable canopy height predictions.

Conversely, *RFH* treats each data point independently and does not incorporate any spatial correlations beyond those already present in the input data. These limitations are apparent in the initial accuracies of *RFH*, which is outperformed by *CHNET* (Table III). This is also particularly evident in the representation of the spatial autocorrelation of *RFH* predictions, which exhibit almost no spatial structure (Figure 4b). This indicates that the model fails to capture the spatial correlations in GEDI reference canopy heights. These spatial correlations can be more complex than simply linking raw input data to canopy heights. In the analysis of their global canopy height model, Lang et al. [9] observed that although there is a clear correlation between classical vegetation indexes like NDVI and the estimated canopy heights, the relationships between image features and canopy height are much more complex. In another study for country-scale canopy height mapping using 10-m resolution S2 images, Lang et al. [72] demonstrated the important role of textural features that correlate with vegetation height. To confirm their hypothesis, they implemented their CNN model by setting the spatial size of all convolution kernels to 1×1, thus forcing it to treat each pixel independently and preventing it from learning any texture or spatial context. In this configuration, the reported errors grew significantly, especially for high canopy heights in the range of 40-60 m. These results demonstrate the benefit of considering textual and spatial features when dealing with high-resolution images, particularly in areas of very high vegetation like French Guiana. These features are not directly leveraged by a technique like Random Forest in the context of this study. As a result, the remaining spatial correlation, which is not accounted for by *RFH*, is instead in the residuals, where the most pronounced spatial structure can be observed (Figure 4b). Contrary to *CHNET* predictions, where some spatial autocorrelation is found, it is entirely contained in the residuals for *RFH*. Implementing a spatial interpolation technique such as RK is therefore even more important for a non-spatial method like *RFH*. Several studies have demonstrated the importance of integrating spatial information for more accurate and reliable predictions. For example, Wang et al. [35] implemented a spatially-weighted geographical Random Forest model which outperformed the traditional Random Forest method, reaching the conclusion that the effects of spatial non-stationarity need to be accounted for in the modeling process.

In our study, the importance of integrating spatial information in canopy height mapping is clear when examining the spatial behaviors of both *CHNET* and *RFH*. The fitted semivariograms of the residuals provide useful insights in this regard. For *RFH*, the residuals show more spatial structure than those of *CHNET* in general. Specifically, the range for *RFH* is larger by more than half a kilometer compared to *CHNET*, indicating that spatial dependencies persist over greater distances for the residuals of *RFH*. This is because *RFH* leaves more spatial correlation unaccounted for in its predictions. Additionally, the semivariogram of *RFH* residuals displays more dynamics, as highlighted by its partial sill (i.e., the difference between sill and nugget) of 3.3 m² compared to 2.0 m² for *CHNET*. This further illustrates how *RFH* residuals contain more spatial structure. Nevertheless, RK provides improvements for both methods. When evaluating the impact of RK, the first observation is that it is effective at unbiasing the results, particularly for *RFH*. Since RK adjusts canopy height predictions based on the residuals, it allows reducing overestimation and underestimation, which are common issues in studies using GEDI data for canopy height mapping in tropical ecosystems. By adding kriged residuals to the initial predictions, underestimated values are increased and overestimated values are reduced, which tends to reduce the bias. Another key aspect of RK corrections is that they are primarily localized around the available GEDI sample points. Conversely, areas lacking GEDI information tend to benefit less from the corrections. As RK is an interpolation method, its effectiveness strongly depends on the density of sample points. Fayad et al. [39] observed in their study using airborne lidar for canopy height mapping that spatial sampling had a significant impact on the results obtained with kriging. Indeed, they found that the accuracy of their canopy height map of French Guiana decreased as the spacing between lidar flight lines increased. Their results proved to be highly sensitive to the spatial sampling of the reference lidar dataset. Similarly in our study, because GEDI measurements are sparse and unevenly distributed, the spatial density of GEDI points is a key factor affecting the overall results obtained with RK. In the end, we demonstrate in this study that there are not enough usable GEDI data points for this method to work effectively at a larger scale. A denser and more uniform distribution would be necessary to achieve more accurate results. A spatial sampling density of at least approximately one GEDI shot every 2 kilometers is recommended. This distance corresponds to the range of spatial autocorrelation observed in our analyses and represents a minimum threshold for effective interpolation. While local areas may currently exhibit sufficient data density, this is not the case at regional scales such as that of French Guiana.

Nonetheless, integrating kriged residuals through RK still offers some improvements in the accuracies of canopy height estimates where sufficient data are available. These improvements are more substantial for *RFH*, as it inherently lacks spatial awareness by itself, but they are still interesting for *CHNET*. Even if CNNs inherently capture some spatial patterns, their architecture alone is not sufficient to fully address the spatial complexity of the data. RK can therefore be a valuable addition, even if it is less impactful than it is for Random Forest. Essentially, when assessed against GEDI data, the combination of *RFH* with RK (*RFH-RK*) produces slightly better results than those obtained from the *CHNET* model alone. This indicates that adding spatial information to Random Forest can bridge the performance gap between a non-spatial method and a convolutional approach that integrates spatial context by design. Other studies have shown that RK can improve the results of Random Forest models for canopy height estimation [39], [73]. This leads to the conclusion that a Random Forest model augmented with spatial information can emulate the capabilities of a convolutional approach in terms of accuracy. This improvement is only possible if the GEDI data are properly distributed, meaning at distances smaller than the range of the spatial autocorrelation and with sufficient spatial density. Kriging is thus a workaround for an imperfect model when it comes to accounting for the spatial autocorrelation in the reference data, but it is only truly effective when there are enough data points. This argues in favor of more usable GEDI measurements than are currently available to achieve truly effective RK for *CHNET* and *RFH*.

V. CONCLUSIONS

This study analyzes and addresses the impacts that sensor-induced anisotropies can have on canopy height estimation using GEDI data, particularly in complex tropical environments like French Guiana. These anisotropies arise from the GEDI sensor's acquisition configuration, mainly from the differing laser beam strengths combined with the spatial sampling of the measurements, which introduces spatial inconsistencies that do not reflect the true spatial autocorrelation of canopy heights. Specifically, the variation in measurements between power and coverage beams, given the fact that power beams offer superior penetration through dense vegetation, leads to discrepancies in the spatial data, which appear as periodic patterns in the residuals of canopy height prediction models.

Ensuring the spatial consistency of canopy height measurements is essential to derive accurate estimates. It is also important to account for the spatial behavior of canopy height, especially since it is a continuous and environment-sensitive variable. Our analyses revealed that refining GEDI reference data to focus exclusively on power beams belonging to a given azimuthal direction, in order to perform the subsequent spatial analysis along that corresponding direction, removed anisotropic patterns and resulted in a more reliable representation of the true spatial autocorrelation of canopy heights. However, a more systematic approach to handling these sensor-related effects from the beginning, possibly through pre-processing techniques, could further enhance the quality of the reference data used to calibrate regression models.

Regarding the integration of spatial correlation for canopy height estimation, our study showed that regression approaches can benefit from spatial interpolation techniques to improve their estimates. Notably, we demonstrated that while U-Net captures part of the spatial correlations, it does not account for all the existing spatial autocorrelation in GEDI measurements. We explored the addition of RK to both *CHNET* and *RFH* models and the results indicated that the incorporation of kriged residuals allowed improving the accuracies of the estimates. These improvements were more significant for the Random Forest algorithm compared to the U-Net architecture. However, for both methods, the corrections were mainly localized around GEDI sample points. The density of available GEDI information appears as a major factor in the effectiveness of spatial interpolation techniques. With the currently available GEDI data, their application at larger scales remains too limited to achieve significant improvements. Finally, our findings suggest that integrating spatial information into non-spatial models like Random Forest can yield results comparable to those achieved by inherently spatial architectures like CNNs. Consequently, a combination of Random Forest with spatial integration methods could serve as an alternative to CNNs for canopy height estimation.

ACKNOWLEDGMENT

This research received funding from the French Space Study Center (CNES, TOSCA 2024 project) and the National Research Institute for Agriculture, Food and the Environment (INRAE). The authors would like to thank the GEDI team and NASA's LP DAAC (Land Processes Distributed Active Archive Center) for proving the GEDI data. The authors also acknowledge Caroline Bedeau (ONF) and Ludovic Villard (CESBIO) for providing additional data used in this study.